\documentclass[12pt,english]{article}
\usepackage[english]{babel}
\usepackage{amsmath,amssymb,amscd,color}
\usepackage{amsfonts}
\usepackage{mathrsfs}
\usepackage{mathtools}
\usepackage{float}
\usepackage{graphicx}
\usepackage{url}
\usepackage{mciteplus}
\usepackage{hyperref}
\usepackage{cite}
\usepackage{physics}
\usepackage{bbold}
\usepackage{braket}
\usepackage[bottom]{footmisc}
\usepackage{verbatim}
\usepackage{paralist}
\usepackage[font=footnotesize,labelfont=bf,width=.9\textwidth]{caption}
\usepackage{multirow}
\usepackage{boldline}
\usepackage{enumitem}
\usepackage{wrapfig}
\usepackage[normalem]{ulem}
\allowdisplaybreaks

\hypersetup{
    colorlinks,%
    citecolor=blue,%
    filecolor=blue,%
    linkcolor=blue,%
    urlcolor=blue
}

\makeatletter
\renewcommand\section{\@startsection {section}{1}{\z@}%
                                   {-5.5ex \@plus -1ex \@minus -.2ex}
                                   {2.3ex \@plus.2ex}%
                                   {\normalfont\large\bfseries}}
\renewcommand\subsection{\@startsection{subsection}{2}{\z@}%
                                     {-3.25ex\@plus -1ex \@minus -.2ex}%
                                     {1.5ex \@plus .2ex}%
                                     {\normalfont\bfseries}}

\numberwithin{equation}{section}

\makeatother

\newcommand{\bea}{\begin{eqnarray}}
\newcommand{\eea}{\end{eqnarray}}
\newcommand{\be}{\begin{equation}}
\newcommand{\ee}{\end{equation}}

\renewcommand{\title}[1]{\vbox{\center\LARGE{#1}}\vspace{5mm}}
\renewcommand{\author}[1]{\vbox{\center#1}\vspace{5mm}}
\newcommand{\address}[1]{\vbox{\center\footnotesize\em#1}}
\newcommand{\email}[1]{\vbox{\center\footnotesize\tt#1}\vspace{5mm}}

\begin{document}

\begin{titlepage}

 \begin{flushright}

\end{flushright}

\begin{center}

\hfill \\
\hfill \\
\vskip 1cm

\title{\Large \bf Gravitational dynamics of near-extreme Kerr (Anti-)de Sitter black holes}

\author{Francesca Mariani$^{a}$, and Chiara Toldo$^{b,c,d}$ 
}

\address{

${}^a$ Department of Physics and Astronomy,
Ghent University, Krijgslaan, 281-S9, 9000 Gent, Belgium
\\
${}^b$ Dipartimento di Fisica, Universita' di Milano, via Celoria 6, 20133 Milano MI, Italy
\\
${}^c$INFN, Sezione di Milano, Via Celoria 16, I-20133 Milano, Italy
\\
${}^d$Physique Th\'eorique et Math\'ematique and International Solvay Institutes, Universit\'e Libre de Bruxelles, C.P. 231, 1050 Brussels, Belgium
}

\email{francesca.mariani@ugent.be, chiara.toldo@unimi.it}

\end{center}

\vfill

\abstract{We analyze the thermodynamic response near extremality of  black holes with angular momentum in $3+1$-dimensional de Sitter and Anti-de Sitter spacetimes. While Kerr-AdS$_4$ is characterized by a single extremal limit, for Kerr-dS$_4$ there are three different extremal scenarios (Cold, Nariai and Ultracold). These exhibit different near horizon geometries, with AdS$_2$, dS$_2$ and Mink$_2$ factors respectively. We analyze each extremal case and contrast the response once the black holes are taken out of extremality. We study the perturbations of the near horizon geometry at the level of the 4D metric, considering a consistent truncation for the metric fluctuations, and find solutions to the linearized Einstein equations. We characterize the perturbations that are responsible for the deviations away from extremality and show that their dynamics is governed by a Schwarzian theory. We treat the Ultracold case separately, detailing how the thermodynamics in 4D is reflected in the near horizon geometry dynamics.
}

\vfill

\end{titlepage}
 \hypersetup{linkcolor=black}
  \tableofcontents
  \hypersetup{linkcolor=blue}

\pagebreak
  \section{Introduction}
  Extremal and near-extremal black holes offer us an important framework to tackle crucial open questions in black hole physics. The former are typically characterized by an enhancement of symmetry near the horizon, which exhibits an AdS$_2$ throat geometry. 
  
  The dynamics of excitations above extremality, characterizing these so-called near-extremal black holes, has been notoriously thorny to study because AdS$_2$ spaces preclude finite energy excitations \cite{Strominger:1998yg,Maldacena:1998uz}. A way forward consists in the introduction of a deformation which breaks the reparametrization symmetries of the AdS$_2$ near horizon geometry. Models of dilaton gravity in 2D, also known as Jackiw-Teitelboim (JT) gravity \cite{Jackiw:1984je,Teitelboim:1983ux} (see \cite{Mertens:2022irh} for a review) capture this by means of a non-trivial profile for the scalar field, breaking the AdS$_2$ conformal symmetry, and displaying at the same time thermodynamic characteristics of black holes, and chaotic behaviour. 
  
The connection between extremal black holes with an AdS$_2$ throat and the JT model has been built in \cite{Maldacena:2016upp, Engelsoy:2016xyb, Almheiri:2016fws, Jensen2016, GrumillerSalzerVassilevich2017,ForsteGolla2017, GrumillerMcNeesSalzerEtAl2017,CadoniCiuluTuveri2018,GonzalezGrumillerSalzer2018} in the context of \textit{near}-AdS$_2$/\textit{near}-CFT$_1$ holography. Several classes of black holes display universal features that can be described by effective 2D gravity theories with dilaton couplings. This is the case of near-extremal Reissner-Nordstr\"om \cite{Nayak:2018qej, Moitra:2018jqs, Sachdev:2019bjn,Castro:2021wzn}, BTZ \cite{Ghosh:2019rcj, Heydeman:2020hhw, Iliesiu:2020qvm,Boruch:2022tno} and Kerr black holes \cite{Castro:2018ffi, Moitra:2019bub, Castro:2019crn, Castro:2021csm, Godet:2020xpk, Castro:2021fhc}, whose thermodynamics response away from extremality
is well known to be described by this $1+1$-dimensional theory \cite{Almheiri:2016fws,Maldacena:2019cbz,Cotler:2019nbi,Moitra:2022glw}.  

This is not an exclusive property of black holes with an AdS$_2$ near horizon geometry. In fact, 
 also the dynamics of other black holes with dS$_2$ and Mink$_2$ factors in their near horizon region is known to be described by 2D dilaton gravity models. This is the specific case of Reissner-Nordstr\"om black holes embedded in $3+1$-dimensional de Sitter space (RNdS$_4$) \cite{Castro:2022cuo}. Working in de Sitter introduces a cosmological horizon, that is known to display a thermodynamic behaviour \cite{PhysRevD.15.2738,Banihashemi:2022htw,Morvan:2022aon}. 
The cosmological horizon of de Sitter makes it possible to have additional extremal limits compared to the case of Reissner-Nordstr\"om black holes in Minkowski. The \textbf{Cold} limit is the usual extremal limit, obtained when the inner and the outer black hole horizons coincide, and is characterized by a near horizon geometry of the form AdS$_2 \times S^2$. The extra cases that arise due to the presence of the cosmological horizon are the so-called \textbf{Nariai} and \textbf{Ultracold} limits \cite{Romans:1991nq,Mann:1995vb,Booth:1998gf}. The Nariai limit is obtained when the outer and cosmological horizons coincide, leading to a dS$_2 \times S^2$ near horizon geometry, while the Ultracold limit occurs when all three horizons (inner, outer, and cosmological) coincide, resulting in a Mink$_2 \times S^2$ geometry. The spherically symmetric sectors of these black holes allow for a dimensional reduction that directly yields the 2D dilaton gravity models of interest, see \cite{Castro:2022cuo} for a detailed analysis.
 
The goal of this paper is to explore deformations of extremal Kerr black holes in $3+1$ dimensional (Anti-)de Sitter space, coined as Kerr-(A)dS$_4$ \cite{Hartman:2008pb, Anninos:2010gh, Anninos:2009yc}, and single out a particular sector of perturbations, responsible for an increase in temperature, that display the Jackiw-Teitelboim dynamics. At the same time, we highlight the differences that arise due to the presence of a non-zero cosmological constant, compared to the case of Kerr black holes in asymptotically flat space.
As RNdS$_4$, Kerr-dS$_4$ admit an inner, an outer and a cosmological horizon whose confluence gives rise to the same three extremal limits. Also these extremal solutions are characterized by the presence of AdS$_2$, dS$_2$ and Mink$_2$ factors in their near horizon geometries respectively. The thermodynamics response of the system to deviations away from extremality is also similar to that of RNdS$_4$, as will be shown later in the paper.

  In this context, we explore axisymmetric gravitational perturbations of the above mentioned extremal solutions at the level of the 4D metric, along the lines of \cite{Castro:2019crn, Castro:2021csm, Godet:2020xpk}, where this approach was followed to investigate the gravitational perturbations of Kerr black holes in $3+1$-dimensional Minkowski space. In that case, one of the modes controlling the departure away from extremality was identified with a JT mode, showing that Kerr also falls in the class of black holes whose near-extremal dynamics is described by JT gravity at the classical level. In the following, we will show that the solutions of the linearized Einstein equations for extreme Kerr-dS$_4$ also admit a scalar mode interpretable as a JT mode. The equations that we find for this mode are compatible with the ones 
 of AdS JT, dS JT and flat JT, depending on which extremal solutions we are deforming. 
Kerr-AdS$_4$ has only one extremal limit whose near-extremal deformations are governed by AdS$_2$ dynamics. Notice that our results for Kerr-AdS$_4$ and cold Kerr-dS$_4$ correctly reduce to the ones of Kerr-Mink$_4$ in the limit where the AdS$_4$ and dS$_4$ curvature radii are sent to infinity ($\ell_{\text{AdS}_4},\ell_{\text{dS}_4}\rightarrow \infty$). We have shown that the departure from extremality is encoded in a field that satisfies the JT-equation \eqref{JT_generic}, whose dynamics is compatible with a Schwarzian action, realizing the breaking of the conformal symmetry of the near horizon region.
 
The paper is organized as follows. In \textbf{Section} \ref{Section2} we review some general features of Kerr black holes in $3+1$-dimensional (Anti-) de Sitter space. We review the three different extremal limits of Kerr-dS$_4$ and characterize their near horizon geometries, showing that contrarily to the AdS and Minkowski cases, here we have extra constraints on the solutions that need to be taken into account. In \textbf{Section} \ref{Section3} we deform the extremal solutions and account for the thermodynamic response of the system to small deformations around extremality, for both Kerr-AdS$_4$ and Kerr-dS$_4$. \textbf{Section} \ref{Section4} is devoted to the analysis of the gravitational perturbations around extremal Kerr-dS$_4$ backgrounds. The formalism we use is inspired by similar procedures used to study the gravitational perturbations around Kerr black holes in $3+1$-dimensional Minkowski space in \cite{Castro:2019crn,Castro:2021csm,Godet:2020xpk}. We motivate the ansatz for the perturbed metric by showing that this naturally emerges as the first order temperature correction to the extremal background. We solve the linearized Einstein's equations, showing that one of the modes used to perturb the background is in fact a JT gravity mode, satisfying the classical JT equations for the dilaton. Finally, in \textbf{Section} \ref{Section5} we perform the same analysis of Kerr-dS$_4$ for Kerr-AdS$_4$. We analyze the extremal solution, that shares many features with the Cold black hole found in de Sitter, and subsequently deform this background by looking at the thermodynamic response of the system to a small increase in temperature. We show that the near-extremal dynamics is captured by a JT mode, and we additionally adopt holographic renormalization and show the appearance of a Schwarzian term in the renormalized on-shell action.

\section{Kerr Black Holes in (Anti-)de Sitter Space}
\label{Section2}
 We analyze rotating black holes in $3+1$-dimensional Anti-de Sitter and de Sitter spacetimes. We start by reviewing their properties and the accessible extremal solutions along the lines of \cite{Hartman:2008pb, Anninos:2010gh, Anninos:2009yc}. In this section, we will mainly focus on the near horizon geometries of the extremal solutions, the definitions of the thermodynamic quantities appearing in the First Law of black hole thermodynamics, and the admitted physical solutions in the phase space, parameterized by $(J,M)$ where $J$ is the angular momentum and $M$ is the physical mass of the black hole. 

Kerr (Anti-)de Sitter black holes are obtained as solutions to Einstein theory with a (negative-)positive cosmological constant $\Lambda$:
\be 
S=\frac{1}{16\pi}\int d^4x~ \sqrt{-g}(R-2\Lambda),
\ee
where $\Lambda$ is related to the curvature radii of AdS$_4$ and dS$_4$ via
\begin{equation}
	\Lambda=\frac{3}{\ell^{2}_{\text{dS}}},\quad \Lambda=-\frac{3}{\ell^{2}_{\text{AdS}}},
	\label{Lambdal}
\end{equation}
respectively. Anti-de Sitter and de Sitter spacetimes are related to each other by the analytic continuation of their curvature radii
\begin{equation}
    \ell_{\text{dS}}=-i\ell_{\text{AdS}}.
    \label{eq:fromdStoAdS}
\end{equation}
From now on, we will refer to both radii with $\ell$, without subscripts. Whether we are referring to dS$_4$ or AdS$_4$ will be clear from the context.

The Kerr-dS$_4$ metric in Boyer-Lindquist coordinates is of the form:
\be
ds^2=-\frac{\Delta_r}{\rho^2}\left(d\tilde{t}-\frac{a}{\Xi}\sin^2\theta d\tilde{\phi}\right)^2+\frac{\rho^2}{\Delta_r}d\tilde{r}^2+\frac{\rho^2}{\Delta_{\theta}}d\theta^2+\frac{\Delta_{\theta}}{\rho^2}\sin^2\theta\left(a d\tilde{t}-\frac{\tilde{r}^2+a^2}{\Xi}d\tilde{\phi}\right)^2,
\label{eq:ds2Kerrds}
\ee
with
\begin{equation}
\label{quant_kerrds}
\begin{split}
&\Delta_r=(\tilde{r}^2+a^2)(1-\frac{\tilde{r}^2}{\ell^2})-2m\tilde{r},\quad \Delta_{\theta}=1+\frac{a^2}{\ell^2}\cos^2\theta\\
&\rho^2=\tilde{r}^2+a^2 \cos^2\theta,\qquad \Xi=1+\frac{a^2}{\ell^2},
\end{split}
\end{equation}
where $a$ is a rotation parameter and $m$ is a mass parameter. The metric of Kerr-AdS$_4$ is obtained from \eqref{eq:ds2Kerrds} after applying the analytic continuation in \eqref{eq:fromdStoAdS}. 

For later convenience, we also write the metric of Kerr-dS$_4$ in Eddington-Finkelstein (EF) or null coordinates:
\begin{equation}
\begin{split}
    ds^2=&-\frac{\Delta_r}{\rho^2}\left(du-\frac{a}{\Xi}\sin^2\theta d\phi\right)^2+2dud\tilde{r}-\frac{2a \sin^2\theta}{\Xi}d\tilde{r}d\phi\\
    &+\frac{\rho^2}{\Delta_{\theta}}d\theta^2+\frac{\Delta_{\theta}}{\rho^2}\sin^2\theta\left(a du-\frac{\tilde{r}^2+a^2}{\Xi}d\phi\right)^2.
    \end{split}
\label{eq:ds2KerrdsNULL}
\end{equation} 
The two forms of the metric \eqref{eq:ds2Kerrds} and \eqref{eq:ds2KerrdsNULL} are related by the coordinate transformation:
\begin{equation}
    d\tilde{t} \rightarrow  du-(\tilde{r}^2+a^2)\frac{d\tilde{r}}{\Delta_r}~,\quad d\tilde{\phi} \rightarrow d\phi-a~\Xi~ \frac{d\tilde{r}}{\Delta_r}~.
\end{equation}
It will be convenient to work with this form of the metric \eqref{eq:ds2KerrdsNULL} when solving Einstein's equations for the perturbations above extremality in later sections.
For later purposes, we also rewrite the metric in \eqref{eq:ds2KerrdsNULL} as\footnote{In Boyer-Lindquist coordinates, the metric can be written as
    \begin{equation}
    \begin{split}
      ds^2= &
   -\frac{\Delta_r\Delta_{\theta}\rho^2}{c(r,\theta,a)}d\tilde{t}^2+\frac{\rho^2}{\Delta(r)}d\tilde{r}^2+\frac{\rho^2}{\Delta_{\theta}}d\theta^2
+\frac{\sin^2\theta c(r,\theta,a)}{\rho^2\Xi^2 }\left(
\frac{\Xi a(\Delta_r-(a^2+r^2)\Delta_{\theta})}{c(r,\theta,a)}d\tilde{t}+ d\tilde{\phi}\right)^2\,.
\label{eq:KerrNulldu4}
    \end{split}
\end{equation}}

\begin{equation}
    \begin{split}
      ds^2= &
   -\frac{\Delta_r\Delta_{\theta}\rho^2}{c(r,\theta,a)}du^2+2dudr-\frac{2a \sin^2\theta}{\Xi}drd\phi+\frac{\rho^2}{\Delta_{\theta}}d\theta^2\\
&+\frac{\sin^2\theta c(r,\theta,a)}{\rho^2\Xi^2 }\left(
\frac{\Xi a(\Delta_r-(a^2+r^2)\Delta_{\theta})}{c(r,\theta,a)}du+ d\phi\right)^2\,,
\label{eq:KerrNulldu4}
    \end{split}
\end{equation}
where we have defined
\begin{equation}
c(r,\theta,a)=(a^2+r^2)^2\Delta_{\theta}-a^2\Delta_r\sin^2\theta\,.
\label{eq:crtheta}
\end{equation}

\subsection{Thermodynamic quantities and phase space}

The physical mass and the angular momentum of a Kerr black hole are defined as 
\begin{equation}
	M=\frac{m}{\Xi^2},\qquad J=Ma.
	\label{Mkerr}
\end{equation}
The entropy, the temperature and the angular velocity at the black hole horizon $r_{h}$ can be expressed in terms of the rotation parameter $a$ as 
 \begin{align}
T_{h}&=\frac{|\Delta_r'(r_h)|}{4\pi r_h^2}
,\\
	\label{tkerr}
	 S_{h}&=\frac{\pi(r_h^{2}+a^{2})}{\Xi},\\
  \Omega_{\text{h}} &=\frac{\Xi a}{r_h^2+a^2}.
 \end{align}
The horizons of the black hole are located at zeros of $\Delta_r$.
For Kerr-dS$_4$ there are four roots of which only one is negative, and the horizons are found at $r=\{r_-,r_+,r_c\}$.
The roots $r_-$ and $r_+$ consist of an inner and an outer black hole horizons respectively, while $r_c$ is the cosmological horizon of de Sitter space and is a direct consequence of having a positive cosmological constant $(\Lambda>0)$. For Kerr-AdS$_4$, $\Delta_r$ has four roots, two of which are negative and therefore non-physical. The two horizons of the black hole are located at $r=\{r_-,r_+\}$.

 We can highlight the different roots of $\Delta_r(r)$ for Kerr-dS$_4$ by rewriting it as
 \begin{equation}
	\Delta(r)=-\frac{(r-r_{c})(r-r_{c_2})(r-r_{+})(r-r_{-})}{\ell^2},
	\label{WfKerr4roots}
\end{equation}
where $r_{c_2}$ is the non-physical negative root.
Comparison of (\ref{quant_kerrds}) and (\ref{WfKerr4roots})
 leads to the following constraints between $a$, $m$, $\ell$ and the roots of the blackening factor:
\begin{equation}
\begin{split}
   & \sum_i r_i = 0,\quad \quad\frac{1}{\ell^2}\prod_i r_i=-a^2,\\
  &(r-r_-)(r-r_+)(r-r_c) =2m\ell^2, \quad  \prod_{i<j} r_i r_j=a^2-\ell^2.
    \end{split}
\end{equation}

\begin{figure}[H]
	\centering
	\includegraphics[width=0.5\linewidth]{"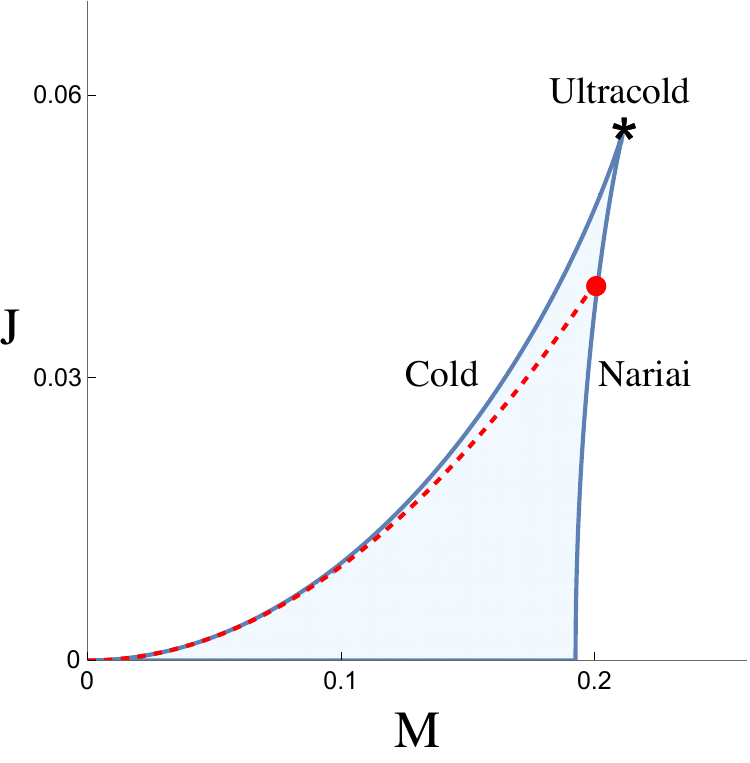"}
	\caption{Shark fin diagram for Kerr-dS$_4$. The value of $\ell$ is fixed to $\ell =1$. The shaded region corresponds to physical solutions, and the white area outside of it corresponds to naked singularities. All the points on the left edge of the diagram are Cold black holes, while all the points on the right correspond to Nariai solutions. The star at the tip of the diagram is the Ultracold point, characterized by the parameters \eqref{eq:uc}. The dashed red line corresponds to the lukewarm branch.}
	\label{SharkFin}
	\end{figure}

The space of admitted physical solutions is very different for Kerr-AdS$_4$ and Kerr-dS$_4$. While in AdS there is no bound on the maximum mass that the black hole can have given a fixed value of angular momentum, the presence of the cosmological horizon of de Sitter space introduces an upper bound on the physical mass of the black hole. Similarly to the Reissner-Nordstr\"om de Sitter case (RNdS$_4$), the space of black hole solutions of Kerr-dS$_4$ consists of a finite region with the characteristic shape of a shark fin, see Figure \eqref{SharkFin}.

\subsection{Extremal limits and near horizon geometries}
\label{subsection:extremal_limits}
Extremal solutions occur when two, or three, horizons coincide. While in AdS$_4$ there is a single extremal case, obtained when the two black hole horizons coincide,
\begin{equation}
    r_-=r_+\equiv r_0,
\end{equation}
in dS$_4$, one has three possible scenarios. This was also the case for non-rotating electrically charged black holes as RNdS$_4$, where the extremal solutions were dubbed as Cold, Nariai and Ultracold. For Kerr-dS$_4$ we have the same scenario, that we recall here:
\begin{itemize}
    \item \textbf{Cold} black holes, obtained when $r_-=r_+\equiv r_0$: correspond to the points on the left edge of the shark-fin. 
    \item \textbf{Nariai} solutions,  $r_+=r_{\text{c}} \equiv r_{\text{n}}$, $r_{\text{n}}>r_{\text{uc}}$: correspond to the points on the right edge of the shark-fin.
    \item \textbf{Ultracold} solutions: $r_-=r_+=r_{\text{c}}\equiv r_{\text{uc}},\quad r_{\text{uc}}=\ell\sqrt{-1+\frac{2}{\sqrt{3}}}$. This is the point at the tip of the diagram and is characterized by
\begin{equation}
  r_{\text{uc}}=\ell\sqrt{-1+\frac{2}{\sqrt{3}}},\quad  m_{\text{uc}}=\frac{4r_{\text{uc}}^3}{\ell^2},\quad a_{\text{uc}}=\frac{\sqrt{3}r_{\text{uc}}^2}{\ell}.
  \label{eq:uc}
\end{equation}
\end{itemize}
One can distinguish between the three different extremal solutions by looking at the polynomial 
\begin{equation}
\Delta(r_{\text{h}})=-1+\frac{6r_{\text{h}}^2}{\ell^2}+\frac{3r_{\text{h}}^4}{\ell^4}
    \begin{cases}
        &< 0~, \quad r_{\text{h}}=r_0,\quad \text{Cold}\\
        &= 0~, \quad r_{\text{h}}=r_{\text{uc}},\hspace{-1.9mm} \quad \text{ultracold}\\
        &> 0~, \quad r_{\text{h}}=r_{\text{n}}, \quad  \text{Nariai}.
    \end{cases}
    \label{eq:polynomial}
\end{equation}
In the following, we will analyze each extremal solution, showing the near horizon geometry near-extremality (NHEK) and the coordinates transformations needed to go from the far away region to near horizon region, $i.e.$ from Kerr to NHEK. 

For completeness, let us mention that for Kerr-dS$_4$ there is another black hole branch called \textit{lukewarm} for which the temperatures at the cosmological and outer horizon coincide $T_+=T_c$, leading to a thermodynamically stable solution \cite{Mellor:1989wc}. This however was not a case of interest of our paper, that focuses only on the extremal limits.

\subsection*{Cold black hole \& extremal AdS black hole}
Both for Kerr-dS$_4$ and Kerr-AdS$_4$, this extremal scenario occurs when the two black hole horizons coincide
\begin{equation}
    r_-=r_+\equiv r_0.
\end{equation}
In de Sitter, this comes with the further constraint $r_0<r_{\text{uc}}$ and the solution is characterized by the extremal parameters:
\begin{equation}
    m_0=\frac{r_0(\ell^2-r_0^2)^2}{\ell^2(\ell^2+r_0^2)},\quad a_0^2=\frac{r_0^2(3r_0^2-\ell^2)}{\ell^2+r_0^2}, \quad \Omega_0=\frac{\Xi a_0 }{(a_0^2+r_0^2)}.
    \label{eq:mextaext}
\end{equation}
The near horizon geometry is found starting from \eqref{eq:ds2Kerrds} by sending 
\be
\tilde{r}\rightarrow r_0+\lambda R_0 r,\qquad \tilde{t}\rightarrow\frac{ tR_0}{\lambda},\qquad \tilde{\phi} \rightarrow \Phi+\Omega_0\frac{tR_0}{\lambda},
\label{eq:NHkerrds}
\ee
where $\lambda$ is an infinitesimal parameter called \textit{decoupling parameter} 
and $R_0$ is a constant defined as 
\be 
R_0=\sqrt{\frac{(r_0+a_0^2)(1+r_0^2 \ell^2)}{1-6r_{0}^2/ \ell^2-3r_0^4/\ell^4}}=\sqrt{-\frac{(r_0+a_0^2)(1+r_0^2 \ell^2)}{\Delta(r_0)}}.
\label{eq:R0C}
\ee
In the second equality we recognized the polynomial of equation \eqref{eq:polynomial}, whose negativity or positivity distinguishes between Cold and Nariai extremal solutions, respectively. For the Cold case, the polynomial in $r_0$ is negative, $\Delta(r_0)<0$, and its negativity correctly ensures a positive argument under the square root.
If we plug (\ref{eq:NHkerrds}) in (\ref{eq:ds2Kerrds}), after taking the decoupling limit  $\left(\lambda\rightarrow 0\right)$ with all the other parameters fixed, we find
\begin{equation}
	\begin{split}
ds^2=\Gamma(\theta)\Big[-r^2dt^2+\frac{dr^2}{r^2}+\alpha(\theta)d\theta^2\Big]+\gamma(\theta)(d\Phi+krdt)^2,
\label{eq:NHEKcold}
	\end{split}	
\end{equation}
where 
\begin{equation}
\Gamma(\theta)=\frac{\rho_0^2R_0^2}{r_0^2+a^2},\quad
	\alpha(\theta)=\frac{r_0^2+a^2}{\Delta_{\theta}R_0^2},\quad
	\gamma(\theta)=\frac{\Delta_{\theta}(r_0^2+a^2)^2\sin^2(\theta)}{\rho_0^2\Xi^2},\quad k=\frac{2ar_0^2\Xi R_0^2}{(r_0^2+a^2)^2},
	\label{eq:gammathetaalpha}
\end{equation}
and $\rho_0$ is just $\rho$ evaluated at the extremal horizon $r_0$. The near horizon geometry is a fibered product of AdS$_2$ and a two-sphere.

The set of diffeomorphisms that preserve the asymptotic metric is \cite{Castro:2019crn}
\begin{equation}
    \begin{split}r&\rightarrow \frac{4r^2f'(t)^2-f''(t)^2}{4rf'(t)^3}\\
    t&\rightarrow f(t)+\frac{2f''(t)f'(t)^2}{4r^2f'(t)^2-f''(t)^2}\\
    \phi&\rightarrow \phi +k ~\text{log}\left(\frac{2rf'(t)-f''(t)}{2rf'(t)+f''(t)}\right),
    \label{eq:diffeoschw}
    \end{split}
\end{equation}
where the function $f(t)$ is a boundary time reparametrization.
Acting with \eqref{eq:diffeoschw} on \eqref{eq:NHEKcold} gives
\begin{equation}
    ds^2=\Gamma(\theta)\Bigg[-r^2\left(1+\frac{\{f(t),t\}}{2r^2}\right)^2dt^2+\frac{dr^2}{r^2}+\alpha(\theta)d\theta^2\Bigg]+\gamma(\theta)\Bigg[d\Phi+kr\left(1-\frac{\{f(t),t\}}{2r^2}\right)dt\Bigg]^2,
    \label{eq:Schw}
\end{equation}
where $\{f(t),t\}$ is the Schwarzian derivative
\begin{equation}
    \{f(t),t\}=\left(\frac{f''}{f'}\right)'-\frac{1}{2}\left(\frac{f''}{f'}\right)^2.
\end{equation}

The extremal Cold near horizon geometry can also be obtained starting from the initial Kerr metric written in null coordinates as in \eqref{eq:ds2KerrdsNULL}, using the same decoupling limit as \eqref{eq:NHkerrds}.
After plugging \eqref{eq:NHkerrds} in the Kerr-dS$_4$ metric in EF coordinates \eqref{eq:ds2KerrdsNULL}, the NH geometry reads
\begin{equation}
    ds^2=\Gamma(\theta)\left[-r^2 du^2+2dudr+\alpha (\theta)d\theta^2\right]+\gamma(\theta)(d\Phi+krdu)^2.
    \label{eq:NHEKcoldEF}
\end{equation}

The coordinate transformation that allows to go from the near horizon geometry in Boyer-Lindquist coordinates \eqref{eq:NHEKcold} to the near horizon geometry in EF coordinates \eqref{eq:NHEKcoldEF} is the following:
\begin{equation}
    dt\rightarrow du-\frac{dr}{r^2},\qquad d\Phi\rightarrow d\Phi+k\frac{dr}{r}.
    \label{eq:transfEFtBL}
\end{equation}

\subsection*{Nariai solution}

This solution occurs in dS$_4$ when the outer black hole horizon coincides with the cosmological horizon, $r_+=r_c\equiv r_{\text{n}}$, with $r_{\text{n}}>0$. The parameters in this case are written in terms of $r_{\text{n}}$:
\begin{equation}
    m_{\text{n}}=\frac{r_{\text{n}}(\ell^2-r_{\text{n}}^2)^2}{\ell^2(\ell^2+r_{\text{n}}^2)},\quad a_{\text{n}}^2=\frac{r_{\text{n}}^2(3r_{\text{n}}^2-\ell^2)}{\ell^2+r_{\text{n}}^2},\quad \Omega_{\text{n}}=\frac{\Xi a_{\text{n}} }{(a_{\text{n}}^2+r_{\text{n}}^2)}.
    \label{eq:aextnariai}
\end{equation}
The near horizon geometry of the Nariai solution is found by sending:
\be
\tilde{r}\rightarrow r_{\text{n}}-\lambda R_0 r,\qquad \tilde{t}\rightarrow t\frac{R_0}{\lambda},\qquad \tilde{\phi} \rightarrow \Phi+\Omega_{\text{n}}t\frac{R_0}{\lambda},
\label{eq:NHkerrdsN}
\ee
where the parameter $R_0$ is defined as follows:
\be 
R_0=\sqrt{\frac{(r_{\text{n}}+a^2)(1+r_{\text{n}}^2/ \ell^2)}{-1+6r_{\text{n}}^2/ \ell^2+3r_{\text{n}}^4/\ell^4}}=\sqrt{\frac{(r_{\text{n}}+a_{\text{n}}^2)(1+r_{\text{n}}^2/ \ell^2)}{\Delta(r_{\text{n}})}}.
\label{eq:R0N}
\ee
In this case the polynomial $\Delta(r_{\text{n}})$ is positive, $\Delta(r_{\text{n}})>0$.
After taking the decoupling limit \eqref{eq:NHkerrdsN} with $(\lambda\rightarrow 0)$ and all the other parameters fixed, the near horizon geometry reads
\begin{equation}
ds^2=\Gamma(\theta)\Big[r^2dt^2-\frac{dr^2}{r^2}+\alpha(\theta)d\theta^2\Big]+\gamma(\theta)(d\Phi+krdt)^2.
\label{eq:NHEKNariai}
\end{equation}
The Nariai NHEK is a fibered product of dS$_2$ and a two-sphere\footnote{Notice that the same geometry can be obtained by taking the same decoupling limit taken in Section 2 of \cite{Anninos:2010gh}, where the relation between their $b$ and our $R_0$ is given by:
\be
R_0^2=\frac{1}{b},\quad b=\frac{r_{\text{n}}(r_{\text{n}}-r_-)(3r_{\text{n}}+r_-)}{\ell^2(a^2+r_{\text{n}}^2)}=\frac{-1+6r_{\text{n}}^2/ \ell^2+3r_{\text{n}}^4/\ell^4}{(r_{\text{n}}+a^2)(1-r_{\text{n}}^2 \ell^2)}.
\ee
}.

\subsection*{Ultracold configuration}
The Ultracold extremal solution is obtained when all the horizons coalesce:
\be 
r_-=r_+=r_c\equiv r_{\text{uc}}\,,
\ee
and the extremal mass and rotation parameters correspond to a single point in phase-space:
\begin{equation}
m_{\text{uc}}=\frac{4r_{\text{uc}}^3}{\ell^2},\quad a_{\text{uc}}=\frac{\sqrt{3}r_{\text{uc}}}{\ell},\quad \Omega_{\text{uc}}=\frac{\Xi a}{(a^{2}+r_{\text{uc}}^2)}.
\end{equation}

The Ultracold near horizon geometry is obtained starting  from the non-extreme Kerr-dS$_4$ black hole \eqref{eq:ds2Kerrds}, where we perform the following coordinate transformation:
\begin{equation}
\begin{split}
    \tilde{r}&\rightarrow r_{\text{uc}}-r_{\text{uc}}\lambda+\sqrt{ \frac{4r_{\text{uc}}^3}{{3r_{\text{uc}}^2+\ell^2}}}\lambda^{3/2}x,\\
     \tilde{t}&\rightarrow \sqrt{\frac{\ell^2+3r_{\text{uc}}^2}{4r_{\text{uc}}}}\frac{t}{\lambda^{3/2}}-\frac{\ell^2}{\sqrt{r_{\text{uc}}}\sqrt{\ell^2+3r_{\text{uc}}^2}}\frac{t}{\sqrt{\lambda}},\\
     \tilde{\phi}&\rightarrow \Phi +\Omega_{\text{uc}}\sqrt{\frac{\ell^2+3r_{\text{uc}}^2}{4r_{\text{uc}}}}\frac{t}{\lambda^{3/2}}.
     \label{eq:uccoordtransf}
    \end{split}
\end{equation}
In the decoupling limit, the 2D metric we obtain is still a metric conformal to Minkowski$_2$.
Starting from the metric of Kerr-dS$_4$ in Eddington-Finkelstein coordinates \eqref{eq:KerrNulldu4} and applying the same coordinates transformation as in \eqref{eq:uccoordtransf}, the near horizon geometry reads:
\begin{equation}
     ds^2=\tilde{\Gamma}(\theta)\left(-dt^2+2dxdt+{\tilde{\alpha}(\theta)d\theta^2}\right)+\gamma(\theta)(d\phi+\tilde{k}xdt)^2,
     \label{eq:ucEF}
\end{equation}
where the functions $\tilde{\Gamma}$, $\tilde{\alpha}$ and $\tilde{k}$ are defined as
\begin{equation}\label{funct_UC}
    \tilde{\Gamma}(\theta)=b\Gamma(\theta)=\frac{\rho_{\text{uc}}^2r_{\text{uc}}}{(r_{\text{uc}}^2+a^2)},\quad \tilde{\alpha}=\frac{\alpha}{b}=\frac{r_{\text{uc}}^2+a^2}{\Delta_{\theta}r_{\text{uc}}
    },\quad \tilde{k}=bk=\frac{2ar_{\text{uc}}^2\Xi}{(r_{\text{uc}}^2+a^2)^2}.
\end{equation}

\vspace{5mm}
To sum up, while AdS$_4$ displays a single extremal solution whose near horizon geometry is a fibered product of AdS$_2$ and a two-sphere, dS$_4$ presents three different near horizon geometries that are fibered products of two-dimensional metrics and a two-sphere, 
\begin{equation}
    ds^2=\Gamma(\theta)(g_{ab}dx^adx^b+\alpha(\theta)d\theta^2)+\gamma(\theta)(d\phi+krdt)^2,
\end{equation}
with $g_{ab}$ being AdS$_2$, dS$_2$ or Mink$_2$.
\section{Thermodynamics and geometry near-extremality}
\label{Section3}
In this section we analyze the thermodynamic response of the system to small deviations away from the three different extremal solutions. While extremal solutions correspond to black hole with coincident horizons, near-extremal solutions are obtained when two (or three) horizons are slightly separated from each other. The displacement of the horizons from the extremal radius is parameterized by the near-extremality parameter $\epsilon$. 
As a consequence of the displacement of the horizons, the system acquires  a small temperature, and the entropy and the angular momentum change accordingly. For near-extremal Kerr-AdS$_4$, Cold and Narai we will however decide to work in an ensemble of fixed angular momentum\footnote{This can be done in gravity by choosing an appropriate boundary counterterm, using results from \cite{Brown:1992bq}.}, $\delta J=0$. The thermodynamic analysis is then very similar to the one of RNdS$_4$, where deviations away from extremality were studied at fixed electric charge, $\delta Q=0$ \cite{Castro:2022cuo}.
\subsection*{Cold and Kerr-AdS$_4$ Black Hole}
The near-extremal Cold black hole is obtained by splitting the inner and the outer black hole horizons as follows
\begin{equation}
    r_-=r_0-R_0\lambda\epsilon+\mathcal{O}(\lambda^2) ,\quad  r_+=r_0+R_0\lambda\epsilon +\mathcal{O}(\lambda^2),
    \label{eq:nearextcold}
\end{equation}
where $R_0$ is defined as in \eqref{eq:R0C}, and $\epsilon$ is a finite parameter that defines the near-extremal solution. 

The near horizon region is reached by sending 
\begin{equation}  \tilde{r}\rightarrow r_0+R_0\left(r+\frac{\epsilon^2}{4r}\right)\lambda+\frac{\epsilon^2\lambda^2}{4r_0},\qquad \tilde{t}\rightarrow \frac{tR_0}{\lambda},\qquad \phi\rightarrow\Phi+\frac{\Omega_0t}{\lambda}.
\end{equation}
Implementing this transformations with $\lambda\rightarrow 0$ holding the other parameters fixed, we find the line element
\begin{equation}
ds^2=\Gamma(\theta)\Big[-r^2\left(1-\frac{4\epsilon^2}{r^2}\right)^2dt^2+\frac{dr^2}{r^2}+\alpha(\theta)d\theta^2\Big]+\gamma(\theta)\left(d\Phi+kr\left(1+\frac{4\epsilon^2}{r^2}\right)dt\right)^2.
\label{eq:NHEKCold}
\end{equation}
This is a \textit{near}-NHEK geometry, which for $\epsilon=0$ reduces to the near horizon geometry of Extreme Cold Kerr \eqref{eq:NHEKcold}.

As a consequence of the displacement of the inner and outer black hole horizons as in \eqref{eq:nearextcold}, the temperature at the outer horizon $r_+$ raises from zero to $T\sim \mathcal{O}(\lambda)$:
\begin{equation}
    T_+=\sqrt{-\frac{\ell^2\Delta(r_0)}{r_0^2(\ell^2-r_0^2)}}\frac{\lambda\epsilon}{2\sqrt{2}\pi} + O(\lambda^2 \epsilon^2).
    \label{eq:temperatureCold}
\end{equation}
The response of the angular momentum to the raise in temperature is  $\delta J \sim \mathcal{O}(\lambda^3)$, and 
the response of the mass as a function of the temperature is:
\begin{equation} \label{Mgap_cold}
    M=M_0+\frac{T_+^2}{M_{\text{gap}}^{\text{Cold}}}+\cdots,
\end{equation}
where $M_0$ is the extremal mass defined starting from \eqref{eq:mextaext}. We identify the mass gap as \begin{equation} M_{\text{gap}}^{\text{Cold}}=-\frac{(\ell^2+3r_0^2)\Delta(r_0)}{4\pi^2r_0^3(r_0^2+\ell^2)},
\end{equation}
 and $\Delta(r_0)$ as in \eqref{eq:polynomial}. Since $\Delta(r_0) <0$ for the Cold black hole, we conclude that the mass gap for the near-extremal Cold black hole is positive, $M_{\text{gap}}^{\text{Cold}}>0$.
The entropy at the outer horizon is 
\begin{equation}
    S_+=S_0+\frac{2T_+}{M_{\text{gap}}^{\text{Cold}}}+\cdots.
\end{equation}
 At the level of the shark-fin diagram, this means that when moving away from extremality, keeping the angular momentum fixed, the mass has to increase.  
$S_0$ is the extremal entropy, $\displaystyle S_0=\frac{2\pi r_0^2\ell^2}{\ell^2+3r_0^2}$. The fact that the mass gap in \eqref{Mgap_cold} is positive is indicative of a positive heat capacity for the Cold black hole.

\subsection*{Nariai solution}
The near-extremal Nariai solution is obtained when the cosmological horizon is separated by a small amount from the outer black hole horizon:
\begin{equation}
    r_+=r_{\text{n}}-R_0\lambda\epsilon,\quad r_{\text{c}}=r_{\text{n}}+R_0\lambda\epsilon,
    \label{eq:nearextN}
\end{equation}
where $R_0$ is defined as in \eqref{eq:R0N} and $\epsilon$ is still a finite parameter.
The near horizon near-extremal Nariai metric is reached by performing the coordinates transformation 
\begin{equation}
    \tilde{r}\rightarrow r_{\text{n}}-\lambda R_0 \left(r-\frac{\epsilon}{2}\right),\qquad \tilde
{t}\rightarrow\frac{R_0}{\lambda} t,\qquad \tilde{\phi}\rightarrow \phi+\frac{\Omega_{\text{n}}R_0}{\lambda}t,
\end{equation}
together with the split of the horizons in \eqref{eq:nearextN}, and reads
\begin{equation}
    ds^2=\Gamma(\theta)\Big[r\left(r-\epsilon\right)dt^2- \frac{dr^2}{r(r-\epsilon)}+\alpha(\theta)d\theta^2\Big]+\gamma(\theta)\left(d\Phi+krdt\right)^2.
    \label{eq:statpatch1}
\end{equation}
From \eqref{eq:statpatch1} we can write the metric of the near-extremal Nariai solution in the static patch by just sending \cite{Anninos:2009yc} 
\begin{equation}
    r\rightarrow\frac{\epsilon}{2}(r+1),\qquad t\rightarrow\frac{2}{\epsilon}t,
\end{equation}
we obtain
\begin{equation}
        ds^2=\Gamma(\theta)\Big[-(1-r^2)dt^2+ \frac{dr^2}{(1-r^2)}+\alpha(\theta)d\theta^2\Big]+\gamma(\theta)\left(d\Phi+krdt\right)^2.
    \label{eq:statpatch2}
\end{equation}
The dS$_2$ part of the metric in \eqref{eq:statpatch2} is written in the static patch coordinates frame, where the radial position of the observer is confined between $r\in (-1,1)$.

As a consequence of the displacement of the horizons, the temperature at the cosmological horizon raises from zero to $T\sim \mathcal{O(\lambda)}$ in this case as well: 
\begin{equation}
T_{\text{c}}=\sqrt{\frac{\ell^2\Delta(r_\text{n})}{r_\text{n}^2(\ell^2-r_\text{n}^2)}}\frac{\lambda\epsilon}{2\sqrt{2}\pi}.
\end{equation}
Keeping the angular momentum fixed up to $\mathcal{O}(\lambda^2)$, the response of the mass is $\delta M \sim\mathcal{O}(\lambda^2)$:
\begin{equation}
    M=M_\text{n}+\frac{T_\text{n}^2}{M_{\text{gap}}^{\text{n}}}+\cdots \qquad \text{with} \qquad
    M_{\text{gap}}^{\text{n}}=-\frac{(\ell^2+3r_\text{n}^2)\Delta(r_\text{n})}{4\pi^2r_\text{n}^3(r_\text{n}^2+\ell^2)}.
\end{equation}
The entropy at the cosmological horizon changes linearly with the 
 temperature:
\begin{equation}
    S_\text{n}=S_c+\frac{2T_c}{M_{\text{gap}}^{\text{n}}}+\cdots
\end{equation}
Since $\Delta(r_\text{n})>0$ for Nariai, the mass gap is negative, $M_{\text{gap}}^{\text{n}}<0$. This means that we can only decrease the mass when we move away from extremal Nariai solutions, consistently with the shape of the diagram in Fig. \eqref{SharkFin}, and the heat capacity at fixed angular momentum is \textit{negative}.

\subsection*{Ultracold black hole}
\label{Section:thermoUC}
The Ultracold black hole represents the most constrained case of the three extremal Kerr-dS$_4$ solutions, displaying a peculiar thermodynamical behavior compared to the Cold and Nariai near-extremal cases.
The first consideration, as already pointed out in \cite{Castro:2022cuo}, is that in this case it is not possible to study deviations away from extremality in the so-called canonical ensemble, where the angular momentum is fixed and the temperature is allowed to increase, as this would take us out of the shark fin diagram. We will instead move to an ensemble where the angular momentum is allowed to change, while the temperature and the angular velocity at the outer horizon, $T_+$ and $\Omega_+$, are fixed\footnote{For this point, we thank M. J. Blacker, A. Castro and W. Sybesma for sharing with us unpublished notes where this decoupling limit in the static case was derived \cite{mattwatse}.}.
The way in which we displace the horizons from the Ultracold radius $r_{\text{uc}}$ is such that the observer lies in the static patch, between $r_+$ and $r_{\text{c}}$.

We choose the following parametrization for $r_-$, $r_+$ and $r_{\text{c}}$: 
\begin{equation}
    r_-=r_{\text{uc}}+\sum_{i=1}^4 m_i\lambda^i+\mathcal{O}(\lambda^5),~  
~ r_+=r_{\text{uc}}+\sum_{i=1}^4 p_i\lambda^i+\mathcal{O}(\lambda^5),~~r_{\text{c}}=r_{\text{uc}}+\sum_{i=1}^4 c_i\lambda^i+\mathcal{O}(\lambda^5),
\label{eq:splitUC}
\end{equation}
where $m_i$'s, $p_i$'s and $c_i'$s are coefficients that are fixed by the requirement that the temperature and the angular velocity don't have corrections up to $\mathcal{O}(\lambda^5)$.
Fixing $T_+$ and $\Omega_+$ up to $\mathcal{O}(\lambda^5)$, means that all coefficients can be basically expressed in terms of $m_1$ only (see Appendix \ref{Appuc} for the expressions of the coefficients). 
For the angular velocity $\Omega_+$ we use the definition given in \cite{Caldarelli:1999xj}, adapted for de Sitter spacetime:
\begin{equation}
    \Omega_+=\Omega_{H}-\frac{a^2}{\ell^2}=\frac{a\left(1-r_+^2/\ell^2\right)}{r_+^2+a^2},
    \label{eq:NewdefOmega}
\end{equation}
where $\Omega_H$ is the angular velocity at the horizon, while the second term in \eqref{eq:NewdefOmega} is simply the angular velocity at $r\rightarrow \infty$, $\Omega_{\infty}$. 
Our strategy here follows that of \cite{mattwatse}. This results in
\begin{equation}
    \Omega_+=-\frac{1}{\ell}+\mathcal{O}(\lambda^5)\,,
\qquad
    T_+=\frac{\sqrt{6+\frac{7\sqrt{3}}{2}}m_1^2\lambda^2}{2\pi\ell^3}+\mathcal{O}(\lambda^5).
    \label{eq:Tuc}
\end{equation}
The first peculiarity is that the temperature scales quadratically with the decoupling parameter, $T_+\sim \mathcal{O}(\lambda^2)$, while in the Cold and in the Nariai cases, as well as in the Kerr-AdS$_4$ black hole, the temperature at the horizon was always scaling linearly with $\lambda$, $T_{\text{h}}\sim \mathcal{O}(\lambda)$.
The entropy, mass and angular momentum scale quadratically with $\lambda$, \textit{i.e.} linearly with the temperature:
\begin{equation}
\begin{split}
    S_+&\sim S_{\text{uc}}+\mathcal{O}(\lambda^2)+\cdots\\
    M_+&\sim M_{\text{uc}}+\mathcal{O}(\lambda^2)+\cdots\\
    J_+&\sim J_{\text{uc}}+\mathcal{O}(\lambda^2)+\cdots,
    \end{split}
\end{equation}
where the ultracold entropy, mass and angular momentum are 
\begin{equation}
     S_{\text{uc}}=\frac{\pi\ell^2r_{\text{uc}}^2(\ell^2+3r_{\text{uc}}^2)}{\ell^4+3r_{\text{uc}}^4},\quad M_{\text{uc}}=\frac{4\ell^6r_{\text{uc}}^3}{(\ell^4+3r_{\text{uc}}^4)^2},\quad J_{\text{uc}}=-\frac{4\sqrt{3}\ell^5r_{\text{uc}}^5}{(\ell^4+3r_{\text{uc}}^4)^2}.
\end{equation}
The main conclusion we can draw at this stage, is that the leading corrections to $M$, $S$ and $J$ are still dictated by the temperature, which scales quadratically with the decoupling parameter. This will have repercussions for the analysis of  the gravitational perturbations around the extremal Ultracold black hole, for which the first order corrections to the near horizon geometry come at $\mathcal{O}(\lambda^2)$ and are dictated by the change in temperature.

To conclude this section, let us mention that  the analysis of the  of the boundary conditions and the charge algebra of Ultracold Kerr-de Sitter black holes was performed in \cite{Detournay:2023zni} and it would be interesting to connect our thermodynamics studies to their findings.

\section{Gravitational perturbations of Kerr-dS$_4$}
\label{Section4}
In this section, we consider gravitational perturbations around the extremal Cold, Nariai and Ultracold background solutions analyzed in section \ref{subsection:extremal_limits}. We will perturb the four-dimensional metric, and show that the linearized Einstein's equations admit as solution a mode that is interpretable as a JT gravity mode.

Our analysis mirrors the one of \cite{Castro:2019crn}, where gravitational perturbations of extreme Kerr black holes in $4$D Minkowski space were considered. Other studies of the gravitational perturbations around extreme Kerr are available for instance in \cite{Castro:2021csm,Godet:2020xpk}. The ansatz that we use here is a simplified one, compared for instance to the one of \cite{Castro:2021csm}, where the perturbations of Kerr are cast using the so called Teukolsky formalism. We choose instead to use simpler perturbations, whose angular dependence is suppressed. While considerably simplifying the treatment, this approach allows us to show that the modes responsible for the near-extremal dynamics are effectively JT gravity modes, leading to Schwarzian dynamics.

In this section, we consider perturbations around Kerr-dS$_4$ and, as done for instance in \cite{Godet:2020xpk}, we find it more convenient to use Eddington-Finkelstein coordinates.
We are interested in capturing the effect of small deviations from the three different extremal solutions and we will use the infinitesimal parameter $\lambda$ introduced in \eqref{eq:NHkerrds} to cast the departure away from the NHEK backgrounds.

\subsection{Setup for the perturbations} \label{setup_pert}

For all three cases, we can write an ansatz for the perturbations in Eddington-Finkelstein coordinates of the following form:
\begin{equation}
   \begin{split}ds^2&=\Big(\Gamma(\theta)+\lambda\chi(u,r)\Big)\left(\kappa r^2  du^2+\lambda\psi(u,r)du^2+\alpha(\theta)d\theta^2\right)+\left(2\Gamma(\theta)+\lambda\eta(u,r)\sin^2\theta\right)dudr\\
&+\Gamma(\theta)\gamma(\theta)\left(\frac{1+\lambda\Phi(u,r)}{\Gamma(\theta)+\lambda\chi(u,r)}\right)\left(d\phi+krdu+\lambda A\right)^2.
 \end{split}
\label{eq:ansatzAll1}
\end{equation}
We parametrize the perturbations in terms of the functions $\chi$, $\psi$, $\eta$, $\Phi$ and a one form $A$ along the $(u,r)$ directions
\begin{equation}
    A=A_u(u,r,\theta)du+A_r(u,r,\theta)dr.
\end{equation}
Some considerations are in order. First, the ansatz for the perturbations in \eqref{eq:ansatzAll1} holds for all three different extremal cases, that are distinguished by different values of $\kappa$:  $\kappa =-1$ for Cold, $\kappa =0$ for Ultracold, $\kappa =1$ for Nariai.

Secondly, the ansatz \eqref{eq:ansatzAll1} is inspired by the form of the the higher order terms obtained from the decoupling limit to the black hole near horizon geometry (see Appendix \ref{pert_full} for a detailed derivation). The reasoning is the following: the (backreaction of the) near-extremal perturbation should build the exterior asymptotic region and this is precisely what the higher order terms of the decoupling limit do, as they are solutions to the Einstein's equations and couple back the near horizon geometry to the full solution. Since we know that the temperature at the outer horizon is linear in the decoupling parameter $\lambda$ \eqref{eq:temperatureCold}, we could suggestively write the ansatz in \eqref{eq:ansatzAll1} for the Cold black hole (and similarly for Nariai and Kerr-AdS$_4$) as
\begin{equation}
  ds^2=\text{NHEK}_{Cold}+T_+\left(\text{finite temperature corrections}\right)+\mathcal{O}(T_+^2).
  \label{eq:Tcorrections}
\end{equation}
 One can show that \eqref{eq:ansatzAll1} fits the decoupling limit of the near horizon geometries of Cold, Nariai and Ultracold black holes when the respective values for $\kappa$ are selected and when the modes $\Phi$, $\chi$, $\psi$ and $\eta$ are constant, independent of time. In Appendix \ref{pert_full} we spell out the first order correction in $\lambda$ to the decoupling limit of the various near horizon geometries.  Since for the Ultracold black hole the temperature at the outer horizon scales quadratically with the decoupling parameter \eqref{eq:Tuc}, the correct ansatz would have perturbations at $\mathcal{O}(\lambda^2)$ instead of $\mathcal{O}(\lambda)$ in \eqref{eq:ansatzAll1}, that would however still result in a perturbation of the metric of the form \eqref{eq:Tcorrections}.
 
In what follows, we will consider a more general family of backgrounds that includes the functions $\mathcal{P}(u)$ and $\mathcal{T}(u)$ \footnote{The metric \eqref{eq:ansatzCold} is obtained from \eqref{eq:ansatzAll1} after applying the coordinates transformation
\begin{equation}
   u\rightarrow\mathcal{F}(u), \quad r\rightarrow \frac{r+\mathcal{G}'(u)}{\mathcal{F}'(u)},\quad \phi\rightarrow \phi-k~\mathcal{G}(u)du,
   \label{eq:transfGF}
\end{equation}
and the functions $\mathcal{P}$ and $\mathcal{T}$ are related to $\mathcal{F}$ and $\mathcal{G}$ through
\begin{equation}
    \mathcal{P}(u)=-2\mathcal{G}'(u)-2\frac{\mathcal{F}''(u)}{\mathcal{F}'(u)},\quad \mathcal{T}(u)=-\mathcal{G}'(u)^2-\frac{2\mathcal{G}'(u)\mathcal{F}''(u)}{\mathcal{F}'(u)}+2\mathcal{G}''(u).
\end{equation}}:
\begin{equation}
\begin{split}ds^2&=\Big(\Gamma(\theta)+\lambda\chi(u,r)\Big)\left(\left(\kappa r^2+r\mathcal{P}(u)+\mathcal{T}(u)\right)du^2+\lambda\psi(u,r)du^2+\alpha(\theta)d\theta^2\right)\\
&+\left(2\Gamma(\theta)+\lambda\eta(u,r)\sin^2\theta\right)dudr+\Gamma(\theta)\gamma(\theta)\left(\frac{1+\lambda\Phi(u,r)}{\Gamma(\theta)+\lambda\chi(u,r)}\right)\left(d\phi+krdu+\lambda A\right)^2,
\label{eq:ansatzCold}
    \end{split}
\end{equation}

In the following, we will solve the Einstein's equations for the metric \eqref{eq:ansatzCold}
\begin{equation}
    R_{\mu\nu}-\Lambda g_{\mu\nu}=0,
    \label{eq:Einstein}
\end{equation}
\begin{equation}
R_{\mu\nu}=R_{\mu\nu}^{(0)}+R_{\mu\nu}^{(\lambda)},\quad g_{\mu\nu}=g_{\mu\nu}^{(0)}+g_{\mu\nu}^{(\lambda)},
\end{equation}
at linear order in $\lambda$, providing the equations for the perturbations parameterized by the fields $\Phi, \chi, \psi, \eta$ and $A$. 

The strategy for solving the linearized Einstein's equations is inspired by the procedure in \cite{Castro:2019crn,Godet:2020xpk}. We start from the $rr$-component of \eqref{eq:Einstein} that gives a linear profile for the dilaton field $\Phi$:
\begin{equation}
    \Phi(u,r)=\phi_0(u)+r\phi_1(u).
    \label{eq:dilaton}
\end{equation}
For all three cases the $\theta\theta$-component of \eqref{eq:Einstein} yields a proportionality relation between the $\chi$ and the $\eta$ fields which reads
\begin{equation}
    \eta=\frac{(\ell^2-3r_0^2)\chi}{2(\ell^2-r_0^2)},
\end{equation}
plus the equation for the $\chi$ field:
\begin{equation} \label{eq_boxchi}
    \Box_2\chi=-2\kappa\chi, 
\end{equation}
where $\Box_2$ is the Laplacian on the AdS$_2$, dS$_2$ and Mink$_2$ metrics, 
\begin{equation}
    g_{ab} dx^a dx^b=(\kappa r^2+r\mathcal{P}(u)+\mathcal{T}(u))du^2+2dudr\,.
    \label{eq:M2EF}
\end{equation}
In particular, we notice that for $\kappa=0$, i.e. for the Ultracold black hole, the $\chi$ field has vanishing Laplacian.
From the $r\theta-$ and $\theta\phi-$ components of \eqref{eq:Einstein} we get $A_r(u,r,\theta)$ and $A_u(u,r,\theta)$, 
From $u\theta$ and $\phi\phi$ we determine the angular component of $A_u(u,r,\theta)$, in this way 
\begin{equation}
\begin{split}
    A&=\alpha+\epsilon_{ab}\partial^a\Psi dx^b,\quad \Psi(u,r,\theta)=\frac{1+\cos^2\theta}{8\Delta_{\theta}\sin^2\theta r_0^2}\left(r_0^2(1+\cos^2\theta)\Phi(u,r)-2\chi(u,r)\right),\\
    \alpha&=\alpha_u(u,r,\theta) du+\alpha_r(u,r) dr, \qquad \alpha_u(u,r,\theta)=a_1(u,r)+a_2(u,\theta),
    \label{eq:Gauge}
\end{split}
\end{equation}
where $\epsilon_{ab}$ is the Levi-Civita tensor defined on the 2D metric \eqref{eq:M2EF}. From the $u\theta-$component we see that $a_2$ only depends on $u$, $a_2(u,\theta)=a_2(u)$.  From the $ur$-equation we determine the equation for the mode $\psi$:
\begin{equation}
\partial_r^2\psi=3\phi_1'+3\left(r-\frac{\mathcal{P}}{2}\right)\phi_1+\frac{2(\ell^4-9r_0^4)}{\ell^2r_0^2(\ell^2-r_0^2)}\chi-\frac{2}{R_0^2}\left((2r-\mathcal{P})\partial_r\chi-2\partial_r\partial_u\chi\right).
\end{equation}
Finally, from the $u\phi-$ and $uu-$ components we can read the JT equations, which are again distinguished by different choices of the parameter $\kappa$:
\begin{equation}
\begin{split}
    \mathcal{P}\phi_1'+\phi_1 \mathcal{P}'-2\kappa\phi_0'-2\phi_1''&=0,\\
   2\mathcal{T}\phi_1'-\mathcal{P}\phi_0'+\phi_1\mathcal{T}'-2\phi_0''&=0.
   \label{eq:JTequations1}
  \end{split}
\end{equation}
The equations in \eqref{eq:JTequations1}, are compatible with the JT equation for the dilaton $\Phi$ on the 2D metrics \eqref{eq:M2EF}
\begin{equation} \label{JT_generic}
   \left( \nabla_a\nabla_b-g_{ab}\Box\right)\Phi-\kappa g_{ab}\Phi=0.
\end{equation}

We have verified that the subleading order in the decoupling limit in App. \ref{pert_full} satisfies the equations above. Notice that for the explicit solution in App. \ref{pert_full} the modes $\chi$, $\Phi$ and $\eta$ are all linear in $r$ and coincide up to a constant multiplicative factor. 

As a recap, from solving the linearized Einstein's equations we are left with the four seemingly independent degrees of freedom: $\Phi, \chi, \alpha_u , \alpha_r $. The first two are determined via the equations \eqref{JT_generic} and \eqref{eq_boxchi} respectively, while $\alpha_u$ and $ \alpha_r $ are undetermined integration functions. In sections \ref{subsection:conicalsing} and \ref{redundancies} we will show that the condition to avoid conical singularities fixes $\chi$ to be proportional to the dilaton field $\Phi$, and we will also find that  $\alpha_u$ and $ \alpha_r $, under certain conditions, can be reabsorbed by a diffeomorphism, therefore are not real degrees of freedom.

\subsection{Conditions from the absence of conical singularities}
\label{subsection:conicalsing}
If we look at the metric of the sphere at fixed radial and time coordinates, we can isolate from \eqref{eq:ansatzCold}
\begin{equation}
ds^2\Big|_{S^2}=\big(\Gamma(\theta)+\lambda\chi(u,r)\big)\alpha(\theta)d\theta^2+\left(1 + \lambda \Phi(u,r) -\frac{\lambda\chi(u,r)}{\Gamma(\theta)}\right)\gamma(\theta)d\phi^2+\mathcal{O}(\lambda^2).
\end{equation}
Given the expression for $\gamma(\theta)$ in \eqref{eq:gammathetaalpha}, the $d\phi^2$ form is ill defined at $\theta=0,\pi$. Expanding around these poles we get to:
\begin{eqnarray}
ds^2\Big|_{S^2}
 &\underset{\theta\rightarrow 0,\pi}{\sim }&
\frac{a_0^2+r_0^2}{\Xi} \left(d\theta^2+\sin^2\theta d\phi^2\right)+  \nonumber \\
&  + & \lambda \frac{\ell^2}{R_0^2} \left(  \chi(u,r) d\theta^2 + (\Gamma (\theta ) \Phi (u,r)-\chi(u,r)) \sin^2 \theta  d\phi^2 \right)  \nonumber \\
& + & \mathcal{O}(\lambda^2).
\end{eqnarray}
It turns out that the conical singularity is then avoided if $\chi(u,r) $ and $\Phi(u,r)$ are proportional to each others, in particular
\begin{equation}
\chi=\frac{\ell^2r_0^2(r_0^2-\ell^2)}{\ell^4-6\ell^2r_0^2-3r_0^4}\Phi
\end{equation}
This relation is compatible with the equations of motion, and shows that the perturbation $\chi$ becomes equivalent to the JT-mode $\Phi$ solving the JT equations \eqref{JT_generic}. 

It is the combination of these two fields that gives rise to the JT dynamics. In the language of \cite{Castro:2021csm}, $\chi$ is a low-lying mode, part of the tower of AdS$_2$ modes in the gauge invariant Weyl scalars appearing in the Teukolsky formulation, and $\Phi$ arises as a result of a compensating, non-single valued diffeomorphism. The combined action of these two gives rise to a smooth perturbation, free of conical singularities at the poles.

\subsection{Redundancies due to gauge freedoms}
\label{redundancies}

In this section we show that the perturbations parameterized by $\alpha_u$ and $\alpha_r$ are pure diffeomorphisms and can be removed by a gauge transformation. We consider the Cold Kerr-AdS$_4$ black hole, but the same analysis applies to the other near-extremal cases as well.

We consider a perturbation of the metric 
\begin{equation}
    \delta g_{\mu\nu}=\mathcal{L}_{\xi} g_{\mu\nu},
    \label{eq:perturbationLie1}
\end{equation}
generated by an infinitesimal diffeomorphisms
\begin{equation} \label{diffeo1}
    \xi^{\mu} (u,r,\theta,\phi) = (c_1 + c_2 u, -c_2 \, r ,0,f_3(u,r))
\end{equation}
where $g_{\mu\nu}$ is the NHEK metric \eqref{eq:NHEKcoldEF}. The action of \eqref{diffeo1} on the NHEK metric \eqref{eq:NHEKcoldEF} gives
\begin{equation}
    \mathcal{L}_{\xi}g=2\gamma(\theta)(\tilde{\alpha}_udu+\alpha_rdr)(d\phi+krdu).
\end{equation}
with 
\begin{eqnarray} \label{f3}
    \alpha_r =  \partial_r f_3(u,r),  \qquad \alpha_u= \partial_u f_3(u,r)
\end{eqnarray}
and $f_3(u,r)$ an arbitrary function of the radial and time coordinates. We hence have shown that if the perturbations $\alpha_r$ and $\alpha_u$ can be written as \eqref{f3}, they can be removed by a diffeomorphism, thereby they are not real physical degrees of freedom.

\subsection*{Aside: Details on gravitational perturbations of the Ultracold solution}
In this portion, we present a more explicit discussion of the perturbations of the Ultracold solution, drawing connections with known results in $\widehat{\text{CGHS}}$ gravity \cite{Callan:1992rs}.

The ansatz for the perturbations around the extremal Ultracold background corresponds to $\kappa=0$ in \eqref{eq:ansatzCold}. 
Following the same procedure as in \ref{setup_pert}, the $rr$ component of the Einstein's equations \eqref{eq:Einstein} still gives a linear profile for the $\Phi$ field \eqref{eq:dilaton}, and the proportionality relation between $\eta$ and $\chi$ is
\begin{equation}
    \eta(u,r)=\frac{1}{4}(3-\sqrt{3})\chi(u,r),
\end{equation}
where we already replaced the value of the Ultracold radius $r_{\text{uc}}$. We then find a vanishing Laplacian for the $\chi$ field, $\Box_2 \chi =0$, and from the conical singularities analysis we know that $\Phi$ is proportional to $\chi$ through $\Phi=\frac{2}{r_{\text{uc}}}\chi$. Upon imposing these constraints we can integrate to get an explicit expression for $\psi$:
\begin{equation}
\psi(u,r)=\psi_0(u)+r\psi_1(u)+\frac{3}{2}r^2b_0\phi_0(u)+\frac{1}{r^2}\left(b_0r+2\mathcal{P}(u)\right)\phi_1(u)+r^2\phi_1'(u),
\end{equation}
where $b_0$ is a constant defined as
\begin{equation}
    b_0=\frac{r_{\text{uc}}(5\ell^4-6\ell^2r_{\text{uc}}^2+9r_{\text{uc}}^4)}{(\ell^3+3\ell r_{\text{uc}}^2)^2}
\end{equation}
The gauge field reads
\begin{equation}
\begin{split}
    A&=\alpha+\epsilon_{ab}\partial^a\Psi dx^b,\quad \Psi=\frac{(\ell^4+3r_{\text{uc}}^4)\left(c_0\Phi(u,r)-2c_1\chi(u,r)\right)}{4\sqrt{3}\ell r_{\text{uc}}\sin^2\theta(\ell^2+3r_{\text{uc}}^2)(2\ell^4+3r_{\text{uc}}^4+3r_{\text{uc}}^4\cos2\theta)},\\
    \alpha&=\alpha_u(u,r,\theta) du+\alpha_r(u,r) dr.
\end{split}
\end{equation}
where $\epsilon_{ab}$ is the Levi-Civita tensor defined on the $2$D metric \eqref{eq:M2EF} with $\kappa=0$. Finally, from $u\phi$ and $uu$, we can read the JT equations
\begin{equation}
\begin{split}
    \mathcal{P}\phi_1'+\phi_1 \mathcal{P}'+2\phi_1''&=0,\\
   -2\mathcal{T}\phi_1'+\mathcal{P}\phi_0'-\phi_1\mathcal{T}'-2\phi_0''&=0.
   \label{eq:JTUC}
  \end{split}
\end{equation}

We notice that the JT equations in \eqref{eq:JTUC}, match precisely with the ones obtained in \cite{Castro:2022cuo} for RNdS$_4$, when considering perturbations around a Mink$_2$ background in the two-dimensional analysis, describing excitations of Ultracold Reissner-Nordstr\"om solutions. In that case, we distinguished between two distinct branches of solutions, and that analysis holds in this case as well. In particular, branch II was defined by constant values of the functions $ \mathcal{P}(u)=\mathcal{P}_0$, $ \mathcal{T}(u)=\mathcal{T}_0$, and admitted the following solutions for $\phi_1$ and $\phi_0$:
\begin{equation}
\begin{split}
    \phi_0(u)&=e^{\mathcal{P}_0u/2}\frac{\mathcal{P}_0}{\mathcal{T}_0}c_1+c_2 e^{-\mathcal{P}_0u/2}+c_3,\\
    \phi_1(u)&=c_0+c_1 e^{\mathcal{P}_0u/2}.
\end{split}
\end{equation}
As pointed out in \cite{Castro:2022cuo}, this solution has the same form as solutions found for models of $\widehat{\text{CGHS}}$ gravity in \cite{Godet:2021cdl}.

\section{Gravitational perturbations of Kerr-AdS$_4$}
\label{Section5}

This section is devoted to the study of the gravitational perturbations around the extremal Kerr-AdS$_4$ background. We work in Boyer-Lindquist coordinates, instead of Eddington-Finkelstein ones. This computation provides a cross-check of our previous analysis, as indeed we find that the resulting perturbations are parameterized in the end by the same physical degrees of freedom. Moreover, later on we will be interested in computing the renormalized on shell action for these perturbations, and we find this procedure is easier done in Boyer-Lindquist coordinates.

\subsection{Perturbations of Kerr-AdS$_4$ in BL coordinates}

Once again we choose an ansatz whose geometry is  similar to the one used in \cite{Castro:2019crn} 
\begin{equation}
    \begin{split}ds^2&=\Big(\Gamma(\theta)+\lambda\chi(t,r)\Big)\left((1+\lambda\psi(t,r))\gamma_{tt}dt^2+\frac{dr^2}{r^2}+\alpha(\theta)d\theta^2\right)\\
&+\Gamma(\theta)\gamma(\theta)\left(\frac{1+\lambda\Phi(t,r)}{\Gamma(\theta)+\lambda\chi(t,r)}\right)\left(d\phi+k a_tdt+\lambda A\right)^2,
\label{eq:AnsatzAdSBL}
 \end{split}
\end{equation}
where $A$ is the usual gauge field with time and radial components and the functions $\gamma_{tt}$ and $a_t$ are defined as
\begin{equation}
  \sqrt{-\gamma_{tt}}=\alpha(t) r+\frac{\beta(t)}{r},\quad a_t=\alpha(t) r-\frac{\beta(t)}{r}\,,
\label{eq:alfabeta}
\end{equation}
while the functions $\Gamma(\theta)$, $\alpha(\theta)$ and $\gamma(\theta)$ are defined for extremal Kerr-AdS$_4$ in \eqref{eq:gammathetaalpha}, adapted to the AdS case by simply choosing the AdS$_4$ radius $\ell_{\text{AdS}_4}$ instead of $\ell_{\text{dS}_4}$. 

The perturbations are now parametrized in terms of the modes $\chi$, $\psi$, $A$ and $\Phi$.
The ansatz in Eddington-Finkelstein coordinates can be obtained after applying the coordinates transformation
\begin{equation}
    dt\rightarrow du+\frac{dr}{r^2}-\lambda \xi\frac{dr}{r^2},\quad d\tilde{\phi}\rightarrow d\Phi-k\frac{dr}{r},\quad \xi= \frac{\ell^2-r_0^2}{r_0(\ell^2+r_0^2)}R_0,
\end{equation}
where the constant $R_0$ is defined for AdS in \eqref{eq:R0C}, plus an additional transformation of the gauge field.

Similar considerations as the ones explained in Section \ref{Section4} hold:
\begin{itemize}
    \item The first contributions to the deformations of the extremal solutions are dictated by the temperature, which at the outer horizon is proportional to the decoupling parameter $\lambda$.
    \item If we start from the Kerr-AdS$_4$ line element in Boyer-Lindquist coordinates as in \eqref{eq:ds2Kerrds} and we apply the coordinates transformation that brings us to the near horizon region \eqref{eq:NHkerrds}, we end up with a line element that matches with \eqref{eq:AnsatzAdSBL} for certain solutions of the modes $\chi$, $\psi$, $\Phi$ and the gauge field $A$, upon a redefinition of the radial coordinate $r$. In particular, one can show that the one-form $A$ is solely supported on the $t-$subspace, $A=A_t(r,\theta)dt$.
\end{itemize}
We proceed now with solving the Einstein's equations. We find the following constraints:
\begin{equation}
 \Phi=\nu(t)r+\frac{\mu(t)}{r},\quad  \beta(t)=\frac{\alpha(t)\mu'(t)}{\nu'(t)} ,\quad\mu(t)=\frac{c_0}{\nu(t)}-\frac{\nu'(t)^2}{4\alpha(t)^2\nu(t)}.
 \label{eq:sources}
\end{equation}
The gauge field $A$ has components
\begin{equation}
    A_t=\alpha_t(t,r)+\frac{\gamma_{tt}(\ell^4+6\ell^2r_0^2-3r_0^4)\left(\Gamma(\theta)\partial_r\Phi-2\partial_r\chi\right)}{2\sqrt{(\ell^2+3r_0^2)(\ell^2-r_0^2)}(\ell^2-3r_0^2)\gamma(\theta)\sin^2\theta\partial_r a_t},
\end{equation}
\begin{equation}
    A_r=\alpha_r(t,r)+\frac{(\ell^4+6\ell^2r_0^2-3r_0^4)\left(\Gamma(\theta)\partial_t\Phi-2\partial_t\chi\right)}{2\sqrt{(\ell^2+3r_0^2)(\ell^2-r_0^2)}(\ell^2-3r_0^2)\gamma(\theta)\sin^2\theta\partial_r a_t}.
    \label{eq:GaugeAdS}
\end{equation}
The field $\chi$ satisfies the equation
\begin{equation}
    \Box\chi=2\chi.
\end{equation}
To avoid conical singularities one needs to impose a similar constraint between the field $\Phi$ and the field $\chi$ as the one imposed in section \ref{subsection:conicalsing}:
\begin{equation}
    \chi=\frac{\ell^2r_0^2(\ell^2+r_0^2)\Phi}{(\ell^4+6\ell^2r_0^2-3r_0^2)}.
    \label{eq:chiphi}
\end{equation}

Also in this case, the perturbations $\alpha_r $ and $\alpha_t$ can be removed by a gauge transformation. We can redefine the one-form $
\alpha$ as
\begin{equation}
    \alpha(t,r)=\alpha_t(t,r) dt+ \alpha_r (t,r)dr, \quad\alpha_t(t,r)=a_1(t,r).
\end{equation}
We moreover implement the following transformation on $\alpha_t$,
\begin{equation}
    \alpha_t\rightarrow -\frac{1}{2}\mathbb{c_0}r^2\partial_r\Phi+\tilde{\alpha}_t,
\end{equation}
with $\displaystyle\mathbb{c}_0=\frac{(\ell^2-3r_0^2)^2\sqrt{\ell^2+3r_0^2}}{2\sqrt{\ell^2-r_0^2}(\ell^2+6\ell^2r_0^2-3r_0^4)}$.
Plugging \eqref{eq:GaugeAdS}-\eqref{eq:chiphi} in the $r\phi$ component of the linearized Einstein's equations, we get the following constraints of the one-form $\alpha$
\begin{equation}
    \partial_r\left(\partial_t\alpha_r-\partial_r\tilde{\alpha_t}\right)=0,
\end{equation}
which is satisfied if $\alpha$ is of the form
\begin{equation}
    \alpha_r=\partial_r F(t,r),\quad\tilde{\alpha}_t=\frac{2(\ell^2-3r_0^2)\sqrt{\ell^2-r_0^2}\sqrt{\ell^2+3r_0^2}\mathbb{c}_3}{\ell^4+6\ell^2r_0^2-3r_0^4}+\partial_t F(t,r),
\end{equation}
where $\mathbb{c}_3$ is a constant coefficient, appearing later in \eqref{eq:Gt}.
We can then choose a vector
\begin{equation}
    \xi=G(t)\partial_t-(rG'(t)+G''(t))\partial_r+F(t,r)\partial_{\phi},
    \label{eq:diffeo}
\end{equation}
where the function $G$ satisfies $G'''(t)=0$,\footnote{The function $G(t)$ is quadratic in time, with coefficients \begin{equation}
    G(t)=\mathbb{c}_1+\mathbb{c}_2 t+\mathbb{c}_3 t^2.
\label{eq:Gt}
\end{equation}} and perform a Lie derivative on the NHEK metric \eqref{eq:NHEKcoldEF} along such vector, obtaining
\begin{equation}
    \delta g_{\mu\nu}=\mathcal{L}_{\xi}g_{\mu\nu} =2\gamma(\theta)(\tilde{\alpha}_tdt+\alpha_rdr)(d\phi+krdu).
\end{equation}
which shows that indeed one can reabsorb the functions  $\alpha_r $ and $\tilde{\alpha}_t$ by choosing $\xi$ to be \eqref{eq:diffeo}.

\vspace{5mm}
\subsection{Boundary time reparametrization \& large diffeomorphisms}
By comparing our ansatz for the perturbations \eqref{eq:AnsatzAdSBL} to the line element obtained after the reparametrization of the boundary time \eqref{eq:diffeoschw}, we see that $\beta$ is induced by a large diffeomorphism, and the two expressions fit if we simply identify $\beta$ with the Schwarzian derivative \cite{Castro:2019vog}
\begin{equation}
    \beta(t)=\frac{\{f(t),t\}}{2},\quad 
    \{f(t),t\}=\left(\frac{f''}{f'}\right)'-\frac{1}{2}\left(\frac{f''}{f'}\right)^2.
    \label{eq:betaSchw}
\end{equation}
The $(t,r)$ sector of the line element obtained in \eqref{eq:Schw} after applying the reparametrization \eqref{eq:diffeoschw} is
\begin{equation}
   g_{ab}dx^adx^b=-r^2\left(1+\frac{\{f(t),t\}}{2r^2}\right)^2dt^2+\frac{dr^2}{r^2}.
   \label{eq:2dbackground}
\end{equation}
The $2$D equations for the scalar field $\Phi$ on the background \eqref{eq:2dbackground} are 
\begin{equation}
    \nabla_a\nabla_b\Phi-g_{ab}\Box\Phi+g_{ab}\Phi=0.
\end{equation}
Given the solution for $\Phi$ in terms of the sources in \eqref{eq:sources}, we immediately see that if we combine \eqref{eq:betaSchw} with the choice $\alpha(t)=1$, we obtain the following relation between the function $\nu$ sourcing the dilaton field $\Phi$ and the diffeomorphism \eqref{eq:diffeoschw}
\begin{equation}
    \left(\frac{1}{f'}\left(\frac{(f'\nu)'}{f'}\right)'\right)'=0.
\end{equation}

The near-extremal near horizon background can be retrieved if the field $\Phi$ is independent of  time and linear in the $r$ variable, which results in $\alpha=1$ and $\beta=-\epsilon^2/4$, and in the following values for the sources
\begin{equation}
   \mu(t)=0,\quad \nu(t)=\text{const},\quad c_0=0.
\end{equation}
This choice also gives a vanishing $r-$component of the gauge field, $A_r=0$. 

\subsection{Holographic renormalization and Schwarzian action}
\label{susection:HolRen}
In this section, we compute the renormalized on shell action for the perturbed Kerr-AdS$_4$ near horizon geometry \eqref{eq:AnsatzAdSBL}, using the results for the perturbations we have obtained in the previous subsection. Given that we deal with a near horizon geometry that contains an Anti-de Sitter factor, in our procedure we follow somewhat closely the standard rules of holographic renormalization in AdS spaces, see for instance \cite{Papadimitriou:2005ii}. 

The renormalized action is obtained by the addition of a counterterm integral, namely
\begin{equation}
    I_{ren}=I_{4\text{D}}+I_{ct},
    \label{eq:Iren}
\end{equation}
where the four-dimensional action  $I_{4\text{D}}$ is the standard Einstein-Hilbert piece plus the Gibbons-Hawking-York boundary term, 
\begin{equation}
    I_{4\text{D}}=\frac{1}{16\pi }\int_M d^4x \sqrt{-g}(R-2\Lambda)+\frac{1}{8\pi}\int_{\partial\mathcal{M}} d^3 x \sqrt{-h}K,
\end{equation}
and the second term in \eqref{eq:Iren} is a counterterm action needed to remove the divergences appearing in $I_{4\text{D}}$ at large $r$, that we choose following the conventions of \cite{Castro:2018ffi,Castro:2019crn}
\begin{equation}
    I_{ct}=\frac{\lambda}{8\pi}\int dt\sqrt{-\gamma_{tt}}~z~\Phi,    \label{eq:ct}
\end{equation}
where $z$ is a coefficient to be determined later.

We will cast our variational problem in terms of the $2$D fields ($\gamma_{tt}$ and $\Phi$), sourced by $\alpha(t)$, $\beta(t)$, $\mu(t)$ and $\nu(t)$, \eqref{eq:sources}. We consider the variation of the $4$D action in terms of the induced metric on the $3$D boundary:
\begin{equation}
\begin{split}
    \delta I_{ren}&=\int_{\partial\mathcal{M}}d^3 x\pi^{\mu\nu}\delta h_{\mu\nu},
    \end{split}
\end{equation}
where $h_{\mu\nu}$ is the induced metric at the boundary $\partial\mathcal{M}$, located at $r\rightarrow \infty$. The variation of the induced metric can be cast in terms of the $2$D fields that are subsequently varied with respect to their sources. After integration over the angular variables $(\theta,\phi)$, the variation of $I_{ren}$ reads 
\begin{equation}
    \delta I_{ren}=\int dt (\pi_{\alpha}\delta\alpha(t)+\pi_{\nu}\delta\nu(t)).
\end{equation}
The $4$D contribution to the momenta $\pi_{\alpha}$ and $\pi_{\beta}$ comes from 
\begin{equation}
\pi^{\mu\nu}_{4\text{D}}=\frac{\delta I_{4\text{D}}}{\delta h_{\mu\nu}}= \sqrt{-h}(K^{\mu\nu}-Kh^{\mu\nu}).
\end{equation}
$\pi_{\alpha,4\text{D}}$ and $\pi_{\nu,4\text{D}}$ both have a divergent $r^2-$term:
\begin{equation}
\pi_{\alpha,4\text{D}}=c_{\mu}\mu(t)+r^2c_{\nu}\nu(t),\quad \pi_{\nu,4\text{D}}=c_{\beta}\beta(t)+r^2c_{\alpha}\alpha(t)
\end{equation}
where $c_{\mu}$, $c_{\nu}$, $c_{\alpha}$ and $c_{\beta}$ are real coefficients satisfying \footnote{The explicit expressions for the coefficients are the following:
\begin{equation}
\begin{split}
    c_{\alpha}&=\frac{\ell^2r_0^2\left(2(\ell^2-3r_0^2)\sqrt{\ell^2+3r_0^2}+\sqrt{\ell^2+r_0^2}(\ell^2+r_0^2)\arctan\left(a_0/r_0\right)\right)}{4\sqrt{\ell^2+3r_0^2}(\ell^4-4\ell^2r_0^2+3r_0^4)},\\
 c_{\beta}&=\frac{\ell^2r_0^2\sqrt{\ell^2-r_0^2}(\ell^2+r_0^2)\arctan{(a_0/r_0)}}{4\sqrt{\ell^2+3r_0^2}\sqrt{\ell^4+2\ell^2r_0^2+3r_0^4}},
    \end{split}
\end{equation}
where $a_0$ is the extremal rotation parameter.}
\begin{equation}
    c_{\mu}=-c_{\beta},\quad c_{\nu}=c_{\alpha}.
\end{equation}
The $r^2-$divergent pieces are correctly removed by the counterterm in \eqref{eq:ct}, if we choose the coefficient $z$ to be
\begin{equation}
    z=-\frac{c_{\nu}}{2}.
\end{equation}
The renormalized action then reads
\begin{equation}
    I_{ren}=\lambda\int dt\Big[\frac{\ell^2r_0^2}{2(\ell^2-r_0^2)}\big(\beta(t)\nu(t)+\alpha(t)\mu(t)\big)+2\alpha(t)\mu(t)c_{\mu}\Big]+\mathcal{O}(\lambda^2)
\label{eq:Iren}
\end{equation}
After using the relation between the source $\beta$ and the Schwarzian derivative in \eqref{eq:betaSchw} and setting the source $\alpha(t)=1$, the renormalized on-shell action (setting $c_0=0$) reads 
\begin{equation}
      I_{ren}=\lambda\int dt\Big[\frac{\ell^2r_0^2}{4(\ell^2-r_0^2)}\{f(t),t\}\nu(t)-\frac{\nu'^2(t)}{4\nu(t)}\left(\frac{\ell^2r_0^2}{2(\ell^2-r_0^2)}+2c_{\mu}\right)\Big]+\mathcal{O}(\lambda^2).
      \label{eq:finalSchwarzian}
\end{equation}
Our final result in \eqref{eq:finalSchwarzian} contains a term with a Schwarzian derivative, where the source $\nu(t)$ plays the role of a time-dependent coupling constant. If we define a new time coordinate as done in \cite{Maldacena:2016upp,Mertens:2019bvy}
\begin{equation}
    d\tilde{t} = \frac{dt}{\nu(t)}\left(-\frac{2}{\ell^2}+\frac{2}{r_0^2}\right),
\end{equation}
we obtain a Schwarzian action with a time-independent coupling constant,
\begin{equation}
      I_{ren}=\frac{1}{2}\lambda\int d\tilde{t}~\{\,f( \tilde{t}),\tilde{t}\,\}-\frac{1}{2}\lambda\int dt~\frac{\nu'^2(t)}{\nu(t)}c_{\mu} .
      \label{eq:finalSchwarzian}
\end{equation}
As pointed out in \cite{Mertens:2019bvy}, in the context of Schwarzian QM, the second term in \eqref{eq:finalSchwarzian} is not physically relevant, leading to an overall factor that does not play any role in the computation of correlators. 

We conclude that the dynamics of our fluctuations are compatible with a Schwarzian effective action, which realizes the breaking of the time reparameterization invariance to $SL(2,\mathbb{R})$, analogously to what happens in Kerr-Minkowski$_4$ \cite{Castro:2019crn}, BTZ \cite{Castro:2019vog} and 5D rotating black holes \cite{Castro:2018ffi}. 

\section{Conclusions and outlook}
In this work we considered the thermodynamic response of Kerr-dS$_4$ and Kerr-AdS$_4$ black holes to deviations away from extremality. We considered a simple perturbation ansatz, eq. \eqref{eq:ansatzCold}, motivated by the $\mathcal{O}(T)$ contributions to the near horizon geometry.
The same formalism has been applied to the three different extremal limits of Kerr-dS$_4$ (Cold, Nariai and Ultracold) and to extreme Kerr-AdS$_4$. The solutions to the linearized Einstein's equations reveal that the combination of the modes $\Phi$ and $\chi$ leads to JT dynamics \eqref{JT_generic}, giving rise to a smooth perturbation, free of conical singularities at the poles. 

By computing the renormalized on-shell action with appropriate counterterms, we also have shown that the equations for these fields are compatible with the Schwarzian effective action \eqref{eq:finalSchwarzian}. While the Cold and Nariai solution can be taken out of extremality by keeping the angular momentum fixed, we have seen that deforming the ultracold solution out of extremality requires working in an ensemble of fixed angular velocity. Investigating more in depth the physical consequences of these ensemble differences is an interesting open direction.

While our work arises from a purely classical perspective, it provides the starting point for the computation of the  quantum corrections to the entropy of de Sitter black holes, along the lines of \cite{Iliesiu:2022onk,Blacker:2025zca,Turiaci:2025xwi,Maulik:2025phe}, whose techniques can be developed also for rotating black holes  \cite{Kapec:2023ruw,Rakic:2023vhv,Kapec:2024zdj,Kolanowski:2024zrq,Arnaudo:2024bbd}. Another possible direction left for exploration is to generalize our simple perturbation ansatz to more general ones with angular dependence, for instance following the analysis of \cite{Castro:2021csm}, making connections to the Newman-Penrose formalism. On a different note, in this paper we also single out a Schwarzian sector for rotating Anti-de Sitter black holes, for which a dual 3D CFT is known. It would be interesting to see this sector in the dual field theory description, see for instance \cite{Cabo-Bizet:2024gny}. We hope to report back on these interesting lines of research in the coming future.

\section*{Acknowledgments}

We thank M. J. Blacker, A. Castro, S. Detournay, T. G. Mertens,  W. Sybesma for useful discussions, comments on the draft and collaborations on related topics. We are particularly grateful to Alejandra Castro for many illuminating discussions, and to M. J. Blacker and W. Sybesma for informing us about the decoupling limit that inspired our procedure in App. \ref{UCapp}.  FM is supported by the Research Foundation - Flanders (FWO) doctoral fellowship 11P9Z24N. The work of CT is supported by the Marie Sklodowska-Curie Global Fellowship (ERC Horizon 2020 Program) SPINBHMICRO-101024314, and the MISU grant 40024018 "Pushing horizons in Black hole Physics". This work was performed in part at Aspen Center for Physics, which is supported by National Science Foundation grant PHY-2210452.

\appendix

\label{Appuc}
\section{Motivating the perturbation ansatz from the decoupling limit \label{pert_full}}

\subsection{Cold/Nariai geometry}
\label{app_nariai_cold}

In this appendix, we show how our ansatz for the perturbed metric \eqref{eq:ansatzCold} is inspired by the form of the corrections to the near horizon geometry coming from higher-order contributions of the decoupling limit of extremal Kerr-dS$_4$ black holes.

More specifically, equation \eqref{eq:ansatzCold} is the result of a meticulous modification of the full Kerr metric in EF coordinates. Each function appearing in the metric is deformed according to the decoupling limit (\eqref{eq:NHkerrds} for Cold and \eqref{eq:NHkerrdsN} for Nariai and \eqref{eq:decLimUC} for Ultracold). For Cold and Nariai, these are linear contributions in the parameter $\lambda$, while for Ultracold, one can show that the contributions grow $\sim$ $\lambda^{2}$, see Section \ref{Section:thermoUC}.  We have checked that the explicit solutions for $\chi, \Phi, \psi, A_u, A_r$ reported in this section indeed solve the perturbation equations we found in sec. \ref{Section4}. We detail here the Cold and Nariai cases, leaving the Ultracold case for the next subsection.

We start by reporting for convenience the Kerr dS metric in EF coordinates, eq. \eqref{eq:KerrNulldu4}:
\begin{equation}
    \begin{split}
      ds^2= &
   -\frac{\Delta_r\Delta_{\theta}\rho^2}{c(r,\theta,a)}du^2+2dudr-\frac{2a \sin^2\theta}{\Xi}drd\phi+\frac{\rho^2}{\Delta_{\theta}}d\theta^2\\
&+\frac{\sin^2\theta c(r,\theta,a)}{\rho^2\Xi^2 }\left(
\frac{\Xi a(\Delta_r-(a^2+r^2)\Delta_{\theta})}{c(r,\theta,a)}du+ d\phi\right)^2\,,
\end{split}
\end{equation}
First of all, the mode $\chi$ arises from the deformation of $\rho^2$, and, if evaluated explicitly, it has the following expression:
\begin{equation}
\chi(r)=-\frac{2R_0r_0(\ell^2+r_0^2)r}{\ell^2\Delta(r_0)},
\end{equation}
with $a_0$ being the extremal rotation parameter, $R_0$ the parameter defined in \eqref{eq:R0C} and $r_0$ the extremal radius. The function $\psi$ is obtained when modifying the term $\displaystyle \frac{\Delta_r\Delta_{\theta}du^2}{c(r,\theta)}$, and is of the form
\begin{equation}
\begin{split}
  \psi(r)&= \frac{4r^3\ell^2r_0(\ell^2-5r_0^2)(\ell^2+r_0^2)^2}{R_0\Delta(r_h)^2}\,.
  \end{split}
\end{equation}
The mode $\Phi$ arises when we apply the decoupling limit to the function $c(r,\theta)$ \eqref{eq:crtheta}: 
\begin{equation}
    \Phi(r)= -\frac{2R_0r_0(\ell^2+r_0^2)r}{r_0^2\ell^2(\ell^2-r_0^2)} \,.
\end{equation}
When we deform the $g_{ur}$ and $g_{r\phi}$ components of the full Kerr metric, we get
\begin{equation}
    2dudr-\frac{2a \sin^2\theta}{\Xi}drd\tilde{\phi}\rightarrow 2\Gamma(\theta)dudr+\lambda A_r(\theta)drd\phi.
\end{equation}
We interpret the function $A_r(\theta)$ as the $r$-component of a gauge field.
From the expansion of the second bracket in \eqref{eq:KerrNulldu4}, we get 
\begin{equation}
    d\phi+ \frac{a(\Delta_r-(a^2+r^2)^2\Delta_{\theta})}{c(r,\theta,a)}du \rightarrow d\phi+krdu+\lambda A_u(r,\theta)
    \label{eq:dudr}
\end{equation} where we interpret $A_u(r,\theta)$ as the $u$-component of the gauge field, which is supported only on the $(t,r)$ subspace: 
\begin{equation}
A(r,\theta)=A_u(r,\theta)du+A_r(\theta)dr.
\end{equation}
Finally, we see the presence of a function $\eta$, responsible of the modification of the $g_{ur}$ term, which explicitly reads:
\begin{equation}
    \eta(r)= -\frac{(\ell^2-3r_0^2) 2R_0r_0(\ell^2+r_0^2)r}{2(\ell^2-r_0^2) \ell^2\Delta(r_0)}\, .
\end{equation}

Summarizing, the ansatz for our perturbed metric can be finally written as:
\begin{equation}
\begin{split}ds^2&=\Big(\Gamma(\theta)+\lambda\chi(u,r)\Big)\left(\kappa (1+\lambda\psi(u,r))du^2+\alpha(\theta)d\theta^2\right)+\left(2\Gamma(\theta)+\lambda\eta(u,r)\sin^2\theta\right)dudr\\
&+\Gamma(\theta)\gamma(\theta)\left(\frac{1+\lambda\Phi(u,r)}{\Gamma(\theta)+\lambda\chi(u,r)}\right)\left(d\phi+krdu+\lambda A\right)^2,
 \end{split}
\label{eq:ansatzAll}
\end{equation}
where $\kappa$ is $-r^2$  for Cold and $r^2$ for Nariai. In Eddington-Finkelstein coordinates, our ansatz \eqref{eq:ansatzAll} differs from the one presented in \cite{Godet:2020xpk}, in the $dudr$ term. We believe that in order to have an ansatz that still fits the near horizon geometry at $\mathcal{O}(\lambda)$, the extra $\sin^2\theta$ term in $g_{ur}$ is needed, also in the flat limit.

\subsection{Ultracold background \label{UCapp}}
In this appendix we report on some details about the near-extremal Ultracold black hole. Taking inspiration from the static case \cite{mattwatse}, we separate the horizons as in \eqref{eq:splitUC},
\begin{equation}
    r_-=r_{\text{uc}}+\sum_{i=1}^4 m_i\lambda^i+\mathcal{O}(\lambda^5),~  
~ r_+=r_{\text{uc}}+\sum_{i=1}^4 p_i\lambda^i+\mathcal{O}(\lambda^5),~~r_{\text{c}}=r_{\text{uc}}+\sum_{i=1}^4 c_i\lambda^i+\mathcal{O}(\lambda^5)\,.
\label{eq:split}
\end{equation}
The  correct coefficients ensuring that $T_+$ and $\Omega_+$ are fixed up to $\mathcal{O}(\lambda^5)$ as in \eqref{eq:Tuc} can be expressed in terms of $m_1$ and $m_4$: 
\begin{equation} \nonumber
        p_1=0,\qquad
        p_2=-\frac{-1+\sqrt{3}m_1^2}{6r_{\text{uc}}},\qquad
        p_3=0,\qquad
        p_4=-\frac{m_1^4}{12\sqrt{3}r_{\text{uc}}^3},
\end{equation}
\begin{equation} \nonumber
        m_2=\frac{(7-\sqrt{3})m_1^2}{24r_{\text{uc}}},\qquad
        m_3=\frac{(23-64\sqrt{3})m_1^3}{384r_{\text{uc}}^2}, \qquad  c_1=-m_1,
\end{equation}
\begin{equation}    
        c_2=-\frac{(-7+4\sqrt{3})m_1^2}{24r_{\text{uc}}},\qquad
        c_3=\frac{(-23+64\sqrt{3})m_1^2}{384r_{\text{uc}}^2},\qquad
        c_4=-m_4+\frac{(18-7\sqrt{3})m_1^4}{72r_{\text{uc}}^3}.
        \label{eq:coefficients}
\end{equation}
The split of the horizons as in \eqref{eq:splitUC} with the choice of the coefficients \eqref{eq:coefficients} ensures that the first order perturbation to the near-extremal Ultracold metric can be written as
\begin{equation}
    ds^2=g_{near-\text{NHEK}_{\mu\nu}}dx^{\mu}dx^{\nu}+\lambda^2h_{\mu\nu}dx^{\mu}dx^{\nu}+\cdots
\end{equation}
where $g_{near-\text{NHEK}}$ is the \textit{near}-extremal metric, obtained as a consequence of the small temperature \eqref{eq:Tuc} turned on by the separation of the horizon as in \eqref{eq:split}. We point out that for $m_1<0$ the ordering of the horizons is preserved ($r_c> r_+ >r_-$). The decoupling limit in this case reads
\begin{equation}
\begin{split}
\tilde{r}&\rightarrow r_{\text{uc}}+\left(\frac{(-2+\sqrt{3})m_1^2}{3r_{\text{uc}}}+\frac{r_{\text{uc}}(c_f+x)}{c_u}\right)\lambda^2,\quad \tilde{u}\rightarrow \frac{c_u}{\lambda^2}u,\\
 \tilde{\phi}&\rightarrow \phi+\frac{c_{\phi}}{\lambda^2}u,\quad c_{\phi}=\frac{\sqrt{-24+14\sqrt{3}}c_u}{r_{\text{uc}}},
\end{split}
\label{eq:decLimUC}
\end{equation}
and yields a line element for the near-extremal background of the form
\begin{equation}
    ds^2=\Gamma(\theta)\big[\left(-1+\mathcal{P}_0r+\mathcal{T}_0\right)du^2+2dudr+\tilde{\alpha}(\theta)d\theta^2\big]+\gamma(\theta)\left(d\phi-\tilde{k}f(x)du\right)^2.
\end{equation}
The coefficients $\mathcal{P}_0$ and $\mathcal{T}_0$ and the function $f$ are compatible with a near-extremal background and are of the form
\begin{equation}
\mathcal{P}_0=\frac{(-3+\sqrt{3})c_um_1^2}{3r_{\text{uc}}^3},\quad \mathcal{T}_0=c_f\mathcal{P}_0-\frac{1}{6}(-3+\sqrt{3})\mathcal{P}_0^2r_{\text{uc}},\quad f(x)=x+c_f.
\end{equation}
The extremal NHEK background is obtained if we send $m_1\rightarrow 0$, together with the coordinate transformation 
\begin{equation}
    r\rightarrow -\frac{x}{\sqrt{2}}+\frac{u}{4\sqrt{2}},\quad u\rightarrow -\sqrt{2}du,\quad \phi\rightarrow d\phi-\frac{\sqrt{3}}{4\ell}du,
\end{equation}
with a subsequent rescaling of $x$, $x\rightarrow x/2$.

We have verified that the perturbations $h_{
\mu\nu}$ for $m_1 =0$ satisfy the equations of motions, and are in form similar to the ones for Cold and Nariai in App. \ref{app_nariai_cold}.

\bibliographystyle{JHEP-2.bst}
\bibliography{all}

\end{document}